\documentstyle[11pt,paspconf,epsf]{article}

\begin{document}

\title{A Mid-Infrared HIRES Atlas of the Galactic Plane}

\author{C. R. Kerton and P. G. Martin}
\affil{Department of Astronomy and CITA, University of Toronto, Toronto, ON, M5S 3H8, Canada}

\begin{abstract} A mid-infrared atlas of the Galactic Plane (MIGA) has been constructed to provide a mid-infrared data set for the Canadian Galactic Plane Survey, a Canadian-lead international project to map all of the major components of the interstellar medium of the Milky Way at a common resolution of approximately $1'$.  The atlas is based on HIRES processing of IRAS 12 and 25 $\mu$m data.  Details of the MIGA production and applications of the atlas are discussed and illustrated. 
 \end{abstract}

\section{Introduction}
        A mid-infrared (12 and 25 $\mu$m) atlas of the Galactic Plane (MIGA) has been constructed using IRAS data processed to approximately $1'$ resolution using the HIRES image construction process (Aumann, Fowler, \& Melnyk 1990).  The new MIGA along with the far-infrared Infrared Galaxy Atlas (IGA) (Cao et al.\ 1996, 1997) are being merged with radio and millimetre data as part of the Canadian Galactic Plane Survey (CGPS), a project to survey about a quadrant of the Galactic Plane at $1'$ resolution over a wide range of wavelengths (e.g., English et al.\ 1998).

        The addition of a mid-infrared data set to this data base is important since one of the goals of the CGPS is to understand the evolution of dust as it moves through different phases of the ISM. The 12 and 25 micron bands of IRAS have been shown to be good tracers of the smallest dust particles (very small grains and large carbonaceous molecules or PAHs) (Onaka et al.\ 1996).  Therefore, the MIGA will allow investigators to study the emission from the smallest components of interstellar dust at an angular resolution that is comparable to that of the complementary data that are being used to define different physical environments.

\section{The Canadian Galactic Plane Survey}

The CGPS is a Canadian-lead international project to map all of the major components of the interstellar medium of the Milky Way at a common resolution of approximately $1'$.  The survey will cover the Galactic Plane from $75^{\circ} < l < 145^{\circ}$ and $-3^{\circ} < b < +5^{\circ}$.  Table 1 summarizes the CGPS data set.

In order to make the survey as useful as possible to astronomers, while not taxing the computing resources of the average user, the survey data will be made available as a series of $5^{\circ}\times 5^{\circ}$ mosaics covering the survey region.  Mosaic construction is being done at the Radio Astronomy Lab at the University of Calgary.  Data archiving and public distribution will be handled by the Canadian Astronomy Data Centre (CADC) in Victoria.  The first mosaic should be released in June 1999.

Currently the CGPS is on target to be completed by March, 2000.  For the latest information about the survey status see the CGPS home page at http://www.ras.ucalgary.ca/CGPS/.  

\begin{table}
\caption{The CGPS Data Base}
\begin{center}\small
\begin{tabular}{ccccc} \tableline
Frequency & Wavelength & Observation & ISM Component & Source \\ \tableline
151 MHz   &  190 cm    & Continuum (I)&  Ionized Gas  &      MRAO   \\  
232 MHz   &  130 cm    &    ''         &    and       &      BAO    \\ 
327 MHz   &   92 cm    &    ''         & Relativistic    &   ''     \\  
408 MHz   &   74 cm    &    ''         &    Plasma  &        DRAO   \\
1420 MHz  &   21 cm    &    ''         &            &        DRAO   \\
    ''  &      ''    & Continuum (QUV) & Magnetic Fields &   ''       \\
    ''      &  ''          & H Spectral Line & Atomic Gas &  ''     \\ 
         &            &               &         &       \\
115 GHz   &   2.6 mm   & CO Spectral Line & Molecular Gas & FCRAO \\ 
         &            &               &         &       \\
 3 THz    &  100 $\mu$m &Continuum (I) &  Large Dust Grains  & IPAC \\ 
 5 THz    &   60 $\mu$m &    ''        &            &      (IGA) \\ 
12 THz    &25 $\mu$m &     ''         &Small Dust Grains &Toronto \\  
25 THz    &12 $\mu$m &     ''          &PAH Molecules     & (MIGA) \\ \tableline \tableline 
\end{tabular}
\end{center}
\end{table}

\section{Processing}

The Mid-Infrared Galaxy Atlas (MIGA) is a mid-infrared counterpart to the far-infrared Infrared Galaxy Atlas (IGA). The basic steps involved in the construction of the MIGA are identical to those used in the construction of the IGA (Cao et al.\ 1997).  Preprocessing steps are done on a Sparc Ultra and YORIC is executed on a SGI Origin 2000.  As with the IGA, the large angular scale preprocessing of the IRAS data allows large high-quality mosaics to be constructed from the final HIRES images.

The MIGA consists of $1^{st}$ and $20^{th}$ iteration HIRES images, along with ancillary maps showing beam shape, coverage, photometric noise and IRAS detector tracks.  A region corresponding to the CGPS survey region in Galactic longitude and  $\pm6^{\circ}$ in latitude has been processed.  Far-infrared images in the CGPS region above the $+4.7^{\circ}$ limit of the IGA are also being produced to provide complete IR coverage of the CGPS region.  Supplementary processing in the immediate future is focussing on select fields in support of the WIRE mission and other projects being undertaken by the CGPS consortium.

A ringing suppression algorithm developed after the IGA started production is being used in the MIGA (Cao, Eggermont, \& Terebey 1997).  Tests on fields processed showed that while the new algorithm does lead to a slight reduction in point-source ringing, the effect is not dramatic.  MIGA images, with and without ringing suppression, were reprojected to the lower-resolution geometry of an ISSA plate and AC/DC corrected.  The MIGA images were then divided by the ISSA plate and log images were formed.  The average $\pm1\sigma$ pixel value in five different $10'$ circular apertures was then measured on each log image.  Little difference was seen between the two MIGA images (see Table 2), and results are similar to those reported for the IGA. Since these tests showed that the images are virtually indistinguishable from regularly processed HIRES images away from point sources, and the algorithm has been shown to be useful in at least one published study (Noriega-Crespo et al.\ 1997), it was adopted for MIGA production.

\begin{table}
\caption{Tests of Ringing Suppression Algorithm}
\begin{center}\small
\begin{tabular}{cccc} \tableline
$l$ & $b$ & Ringing Sup. vs. ISSA & Regular vs. ISSA   \\ \tableline
142.2  & $-1.52$      & $-0.0015\pm0.034$   &  $-0.0018\pm0.034$  \\
142.4  & $-1.95$      & $0.0001\pm0.034$    &  $-0.0001\pm0.034$  \\
142.2  & $-2.41$      & $-0.0033\pm0.037$   &  $-0.0039\pm0.036$  \\
141.5  & $-2.43$      & $-0.0026\pm0.035$   &  $-0.0024\pm0.034$  \\
141.5  & $-1.81$      & $-0.0028\pm0.036$   &  $-0.0029\pm0.036$  \\ \tableline \tableline 
\end{tabular}
\end{center}
\end{table}
 \begin{figure}
\plotone{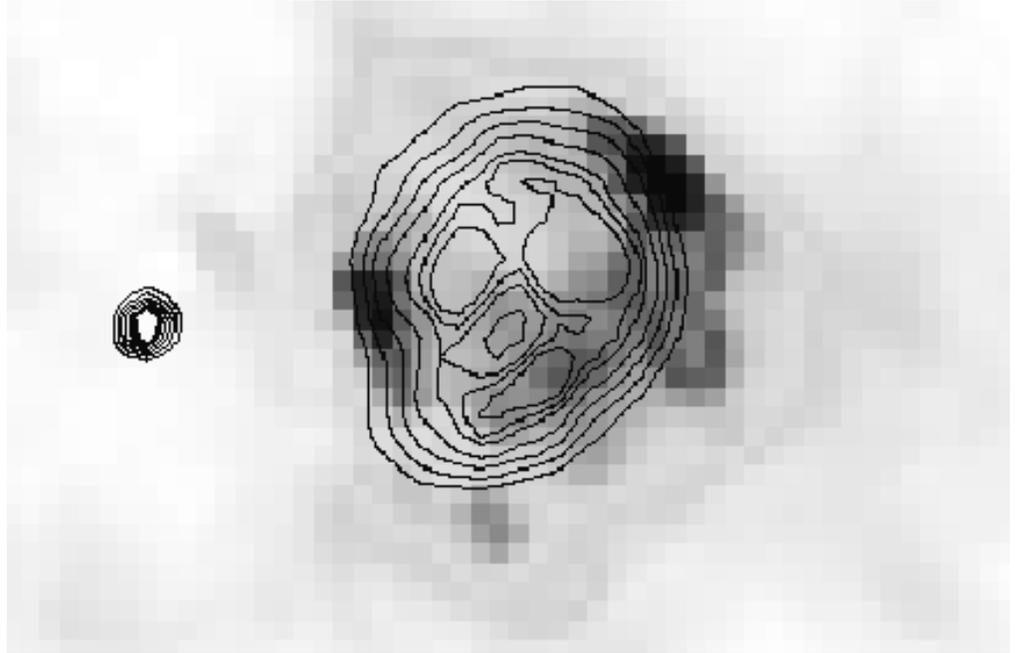}
\caption{This $18'\times10'$ image shows 12$\mu$m emission (greyscale) and 1420 MHz continuum emission (contours) from the small HII region KR 140.  See text for details.}
\end{figure}

The effective beam shape of a HIRES image is a complex function of wavelength, scan geometry, and coverage of a region. This makes the construction of high-resolution ratio maps problematic. If the beam is roughly Gaussian, and does not vary much over the field, then the data can be simply convolved to a common resolution. For more complicated cases, a couple of techniques (cross-scan simulation and prior-knowledge input) have been developed (Fowler and Aumann 1994); however, these are not readily available to the average user of the MIGA and IGA data sets since they require the use of the YORIC program. Common-resolution mosaics of regions of interest to the CGPS consortium will be constructed using the cross-scan simulation method in the future. We are currently investigating the effectiveness of the various techniques on MIGA images with both simple and complex beam shapes.

\section{Multi-wavelength Studies of the ISM}
        The MIGA will be most useful when combined with data taken at other wavelengths as part of the CGPS. Our study of the HII region KR 140 (Figure 1) is
a good example of the power of multi-wavelength studies.  KR 140 is a small ($8.5'$ diameter, $\sim$6 pc), isolated HII region at $l=133.4^{\circ}$ $b=+0.1^{\circ}$.  Located in the W3/W4/W5 region, it was chosen for detailed study because of its apparent simplicity --- small, spherical, and excited by a single star (Ballantyne, Martin, \& Kerton 1999; Kerton, Ballantyne, \& Martin 1999).  The image in Figure 1 has contours of 1420 MHz continuum emission over a 12$\mu$m HIRES image convolved to the 1$'$ radio resolution.  It shows the PAH emission peaking in the photodissociation region surrounding the HII region. This is in contrast to HIRES images at 25$\mu$m which show dust emission extending into the ionized gas, consistent with the idea that PAHs are being destroyed by the UV radiation from the central O star. Radio continuum and HIRES images such as these are being used to study the effect of UV radiation on dust. HII regions like KR 140 are planned targets for the WIRE AI program in which we are involved, and we plan to investigate this phenomenon in more detail using the higher-resolution WIRE data.

\end{document}